\documentclass[11pt]{article}
\usepackage{amsmath}
\usepackage{amssymb}
\usepackage{amsfonts}
\usepackage{amscd}
\usepackage{accents,cite}
\usepackage{bbm} 
\usepackage{pst-all}
\usepackage{epsfig}
\date{\today}
\newcommand{\bmat}{\left(\begin{array}}
\newcommand{\emat}{\end{array}\right)}
\newcommand{\be}{\begin{equation}}
\newcommand{\ee}{\end{equation}}
\newcommand{\ba}{\begin{eqnarray}}
\newcommand{\ea}{\end{eqnarray}}

\def\lsim{\raise0.3ex\hbox{$\;<$\kern-0.75em\raise-1.1ex\hbox{$\sim\;$}}}
\def\gsim{\raise0.3ex\hbox{$\;>$\kern-0.75em\raise-1.1ex\hbox{$\sim\;$}}}

\def\be{\beta}

\begin{document}
\vspace*{-.6in} \thispagestyle{empty}
\begin{flushright}
ADP 12-28-T795
\end{flushright}
\begin{flushright}
DESY 12-099
\end{flushright}

\baselineskip = 20pt

\vspace{.99in} {\Large
\begin{center}
{\bf First LHC Constraints on Neutralinos  }

\end{center}}

\vspace{.5in}

\begin{center}
{\bf  Herbi K. Dreiner$^1$, Jong Soo Kim$^2$ and Oleg Lebedev$^3$}  \\

\vspace{.5in}

\emph{$^1$Physikalisches Institut \& Bethe Center for Theoretical Physics, Nu{\ss}allee 12, 53115 Bonn, Germany}

\emph{$^2$  ARC Centre of Excellence for Particle Physics at the Terascale,
School of Chemistry and Physics, University of Adelaide, Adelaide, Australia  }

\emph{$^3$DESY Theory Group, 
Notkestrasse 85, D-22607 Hamburg, Germany
 }
\end{center}

\vspace{.5in}

\begin{abstract}
  The ATLAS and CMS collaborations have recently reported tantalizing
  hints of the existence of a 125 GeV Higgs--like particle, whose
  couplings appear to match well the Standard Model (SM)
  expectations. In this work, we study implications of this
  observation for the neutralino sector of supersymmetric models,
assuming that the Higgs signal gets confirmed. In
  general, the Higgs decay into neutralinos can be one of its dominant
  decay channels. Since a large invisible Higgs decay branching ratio
  would be in conflict with the data, this possibility is now
  constrained.  In particular, we find that most of the region $\mu <
  170$ GeV, $M_1 <70$ GeV at $\tan\beta\sim 10$ and $\mu < 120$ GeV,
  $M_1 <70$ GeV at $\tan\beta\sim 40$ is disfavored.
\end{abstract}

\newpage 

\section{Introduction}

The LHC experiments have given a possible first indication of the
Higgs boson at a mass around 125 GeV
\cite{Chatrchyan:2012tw,Chatrchyan:2012tx,Chatrchyan:2012ty,Chatrchyan:2012dg,ATLAS:2012ae,ATLAS:2012ad,ATLAS:2012ac}. 
The main production mechanism in the Standard Model (SM) 
is gluon fusion $gg\to h$
\cite{Georgi:1977gs}. At the subleading level,  vector boson fusion  $qq\to qq h$   
also contributes  \cite{Cahn:1983ip}.
The \texttt{CMS} and
\texttt{ATLAS} searches are based on several decay channels of the
Higgs: $h\to\gamma\gamma$ \cite{Ellis:1975ap}, $h\to W^+W^-$
\cite{Rizzo:1980gz,Fleischer:1980ub,Dittmar:1996ss}, and $h\to ZZ$ 
\cite{Fleischer:1980ub}. The dominant decay mode of a 125 GeV mass 
Higgs is $h\to b\bar b$, for which the background is however too
large.  In this paper we are interested in a potential invisible decay
width of the Higgs boson. The total decay width of the SM Higgs 
 is about $\Gamma_h\approx4.2
\,$MeV for a Higgs mass of 125 GeV \cite{Bechtle:2011sb}. This is below the resolution
of the LHC and can thus not be directly measured in the resonance
channels $h\to\gamma\gamma$ and $h\to ZZ$, where the final state can
be reconstructed. A discrepancy from the theoretical value for the total
width would be a direct indication of additional contributions beyond
the SM. All the same, in a given production and decay channel, the
event rate is proportional to the production cross section times the
decay branching ratio, \textit{e.g.}
\begin{equation}
\mathrm{Rate}_{\gamma\gamma}= \sigma(pp\to h+X)\times \mathrm{BR}(h\to
\gamma\gamma) \times {\cal{L}},
\end{equation}
where $X$ depends on the production mechanism and
$\cal{L}$ is the luminosity. Thus, via the branching ratio the
total width enters indirectly in the event rate.  If we take a given
model, for example the SM, and extend it by adding a hypothetical
invisible decay width to the Higgs boson as a free parameter $\Gamma_
{\mathrm{inv}}=\Gamma(h\to\mathrm{inv.})$, we can perform a fit
of $\Gamma_{\mathrm{inv}}$ to the observed event rates, assuming the
Higgs mass and the SM Higgs production mechanisms. Two such global
fits have recently been performed in $(a)$ Ref.~\cite{Giardino:2012ww}
and $(b)$ Ref.~\cite{Espinosa-inv}, resulting in the upper bounds
\begin{eqnarray}
&(a)& {\rm BR_{inv}} < 0.15 \;(0.30) \;\;\label{bound1}\\
&(b)& {\rm BR_{inv}} < 0.37 \;(0.69) \;\;\label{bound2}
\end{eqnarray}
at 68\% (95\%) CL (see also \cite{Barger:2012hv}). 
As the statistics are not sufficient to claim the
Higgs boson discovery, these constraints should be interpreted with
caution. Nevertheless, one may already explore implications of these
results for new physics.  For example, the bounds on the invisible
Higgs decay set rather strict constraints on Higgs--portal dark matter
models \cite{Djouadi:2011aa} where ${\rm BR_{inv}}$ can be as large as
80\% or more \cite{Lebedev:2011iq}. Early work on invisible Higgs
decays in minimal extensions of the SM also employed other Higgs
production mechnaisms: $tth$ Higgs strahlung \cite{Gunion:1993jf},
associated $Zh$ or $Wh$ production
\cite{Choudhury:1993hv},\cite{Frederiksen:1994me}, and in
Ref.~\cite{Eboli:2000ze} vector boson fusion.

Here we wish to explore the implications of the constraints in
Eqs.~(\ref{bound1}),\,(\ref{bound2}) for the minimal supersymmetric
standard model (MSSM) \cite{Haber:1984rc} and, in particular, for the
neutralino sector thereof. Due to LEP, Tevatron and LHC constraints it
is clear that if supersymmetry exists, most of the superpartners are
heavy, \textit{i.e.} well above the purported Higgs mass scale.
However, it is well known, that there is no lower limit on the mass of
the lightest neutralino
\cite{Choudhury:1999tn,Dreiner:2003wh,Dreiner:2009ic,Dreiner:2011fp}. 
Therefore, the
invisible decay of the Higgs boson to two neutralinos is open and can
even be dominant. In the next section we discuss the Higgs decay to
neutralinos and the constraints on the supersymmetric parameter space
resulting from Eqs.~(\ref{bound1}),\,(\ref{bound2}). In Sect. 3 we
conclude.

\section{Higgs decay into neutralinos  }

The Higgs decay into neutralinos has been studied in
Refs.~\cite{Griest:1987qv,Djouadi:1992pu,Djouadi:1999xd,Belanger:2001am}
(see also \cite{Vasquez:2012hn,Desai:2012qy}). In general, it can be
the dominant decay channel if kinematically allowed.  The main
constraint on this scenario comes from the invisible Z--decay, which
has been measured very precisely. However, the uncertainty in the
invisible Z--decay width $\Delta \Gamma_Z^{\rm inv}=\mathcal{O}(1\,\mathrm{MeV})$ is comparable to
the total SM Higgs width $\Gamma_h$,
\begin{equation}
\Delta \Gamma_Z^{\rm inv} \sim \Gamma_h \;.
\end{equation}
Therefore, ${\cal O}(1)$ invisible Higgs decay branching ratio can be compatible 
with the Z--pole data. (Further constraints are imposed if the neutralino
is assumed to be thermal dark matter \cite{Belanger:2001am}).

To make our analysis  more transparent,
we will assume that the sfermions, gluinos and charged Higgses
are sufficiently heavy (TeV--scale)  so that the production cross
section for the lightest Higgs $h$   is SM--like. This is certainly consistent
with (and perhaps hinted by) the current LHC bounds on superpartners 
(see also \cite{Heinemeyer:2011aa}).
Specifically, in terms of the FeynHiggs  \cite{Heinemeyer:1998yj}   variables, 
we choose $M_{\rm SUSY}=M_A=1$ TeV and adjust $A_t$ for a given $\tan\beta$ 
to obtain $m_h=125 \pm 1$ GeV. We use the  FeynHiggs version 2.8.6 with the 
default settings and $m_t=172$ GeV.

The Higgs decay width into the lightest neutralinos $\chi^0_1$ is given by
\cite{Griest:1987qv}
\begin{equation}
\Gamma (h \rightarrow \chi^0_1 \chi^0_1)= \frac{G_F M_W^2 m_h}{2 \sqrt{2} \pi}~
\left(  1- 4 m_{\chi^0_1 }^2/m_h^2   \right)^{3/2} 
\big\vert  C_{h \chi^0_1 \chi^0_1 }   \big\vert^2 \;,
\end{equation}
with 
\begin{equation}
 C_{h \chi^0_1 \chi^0_1 } = \big( N_{12} -\tan \theta_W \; N_{11}   \big)
\big( \sin\beta \; N_{14} -\cos\beta \; N_{13}   \big)\;.
\end{equation}
Here $\tan\beta = \langle H_2^0 \rangle /  \langle H_1^0 \rangle$ and 
$N_{ij}$ is the orthogonal\footnote{We assume CP--conserving soft terms.}   matrix which 
diagonalizes the neutralino mass matrix  \cite{Haber:1984rc}:
\begin{equation}
N~M_{\chi^0}~N^T = {\rm diag} \; (m_{\chi^0_1}, m_{\chi^0_2},m_{\chi^0_3},m_{\chi^0_4} )
\end{equation}
with 
\begin{equation}
M_{\chi^0} =  \left(
\begin{matrix}
M_1 & 0 & -M_Z \sin\theta_W \cos\beta & M_Z \sin\theta_W \sin\beta \\
0 & M_2 & M_Z \cos\theta_W \cos\beta & - M_Z \cos\theta_W \sin\beta \\
-M_Z \sin\theta_W \cos\beta & M_Z \cos\theta_W \cos\beta & 0 & -\mu \\
M_Z \sin\theta_W \sin\beta &  - M_Z \cos\theta_W \sin\beta & -\mu & 0 
\end{matrix}
\right) \;.
\end{equation}
The analogous Z--width
 is given by \cite{Heinemeyer:2007bw}
\begin{equation}
\Gamma (Z \rightarrow \chi^0_1 \chi^0_1)=
\frac{\alpha}{3} M_Z \; \left(  1- 4 m_{\chi^0_1 }^2/M_Z^2   \right)^{3/2} \;
\big\vert  C_{Z \chi^0_1 \chi^0_1 }   \big\vert^2 \;,
\end{equation}
where 
\begin{equation}
C_{Z \chi^0_1 \chi^0_1 } = \frac{1}{2 \cos\theta_W \; \sin \theta_W}\;  \big( N_{14}^2- N_{13}^2 
\big) \;.
\end{equation}
The relevant LEP constraint is \cite{ALEPH:2005ab}
\begin{equation}
\Gamma (Z \rightarrow \chi^0_1 \chi^0_1) < 3 ~{\rm MeV}
\end{equation}
at 95\% CL.
We observe that both the Higgs and Z decay 
rates involve couplings to the Higgsino components of the 
neutralino  $N_{13}, N_{14}$   and as such   vanish in the pure bino limit.
For typical values of $\tan\beta \sim 10$, the Higgs decay is controlled by the $H_2$ 
Higgsino component $N_{14}$, whereas the Z decay involves both  $N_{13}$ and $ N_{14}$.
As the bino mass $M_1$ decreases, $N_{14}$ becomes small whereas $N_{13}$ 
remains substantial\footnote{Unlike $N_{14}$, $N_{13}$ does not vanish as 
$M_1 \rightarrow 0$, $\cos\beta \rightarrow 0$. This limit corresponds to the 
massless  bino--$H_1$-Higgsino  $\chi_1^0$. }. 
In this limit, the Z--width imposes a strict constraint.
On the other hand, for higher $M_1$ and especially above the kinematic limit 
for  $Z \rightarrow \chi^0_1 \chi^0_1 $,    the Higgs invisible width 
can be  comparable to the SM Higgs width without violating the Z--bound.
Here we treat  $M_1$ and $M_2$ as free parameters and do not impose
the supersymmetric grand unified theory constraint $M_1=(5/3) \tan^2\theta_W M_2$.
Therefore, the stricter PDG bound $m_{\chi^0_1}>46\,$GeV
\cite{Nakamura:2010zzi} does not apply.

The other relevant collider constraints are imposed by the chargino mass
bound 
\begin{equation}
m_{\chi^+} > 94 ~ {\rm GeV}
\end{equation}
and the LEP bound on the neutralino production \cite{Abbiendi:2003sc}
\begin{equation}
\sigma(e^+ e^- \rightarrow \chi_1^0 \chi_2^0) \times {\rm BR}(\chi_2^0 \rightarrow
q \bar q \chi_1^0 ) < 50~{\rm fb} \;. \label{sigma}
\end{equation}
The dominant neutralino production mechanism is due to the
$t$--channel slepton exchange \cite{Ellis:1983er}. This is however
strongly suppressed for slepton masses close to 1 TeV. The
$s$--channel production mediated by the Z--boson is insignificant and,
in the parameter region of interest,
we find that the constraint (\ref{sigma}) is never violated  
once the other bounds are satisfied. A similar conclusion was reached in
\cite{Dreiner:2006sb,Dreiner:2009ic}.

\begin{figure}
\hspace*{0.5cm}
\includegraphics[width=6.9cm]{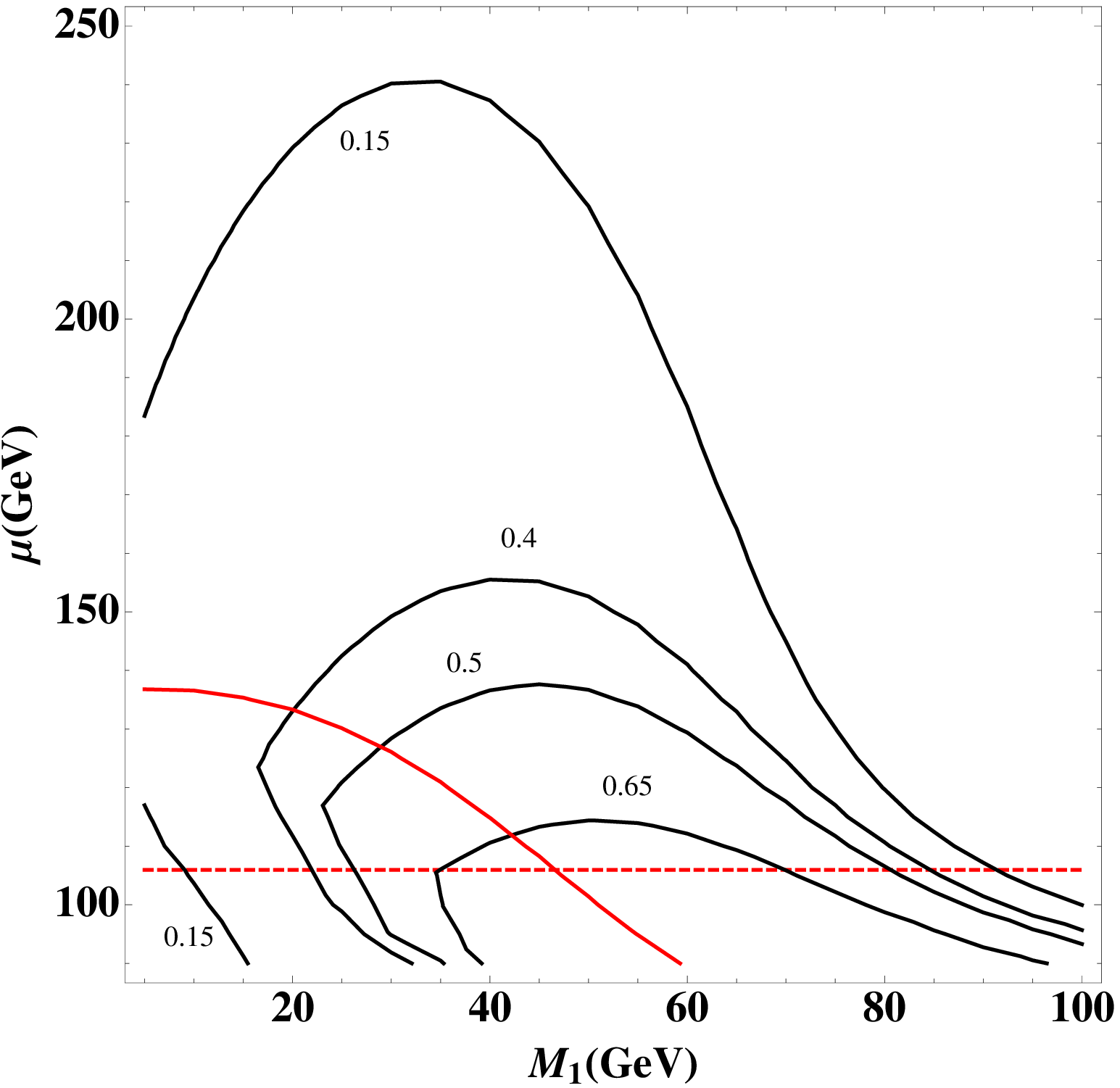}
\hspace*{0.4cm}
\includegraphics[width=6.9cm]{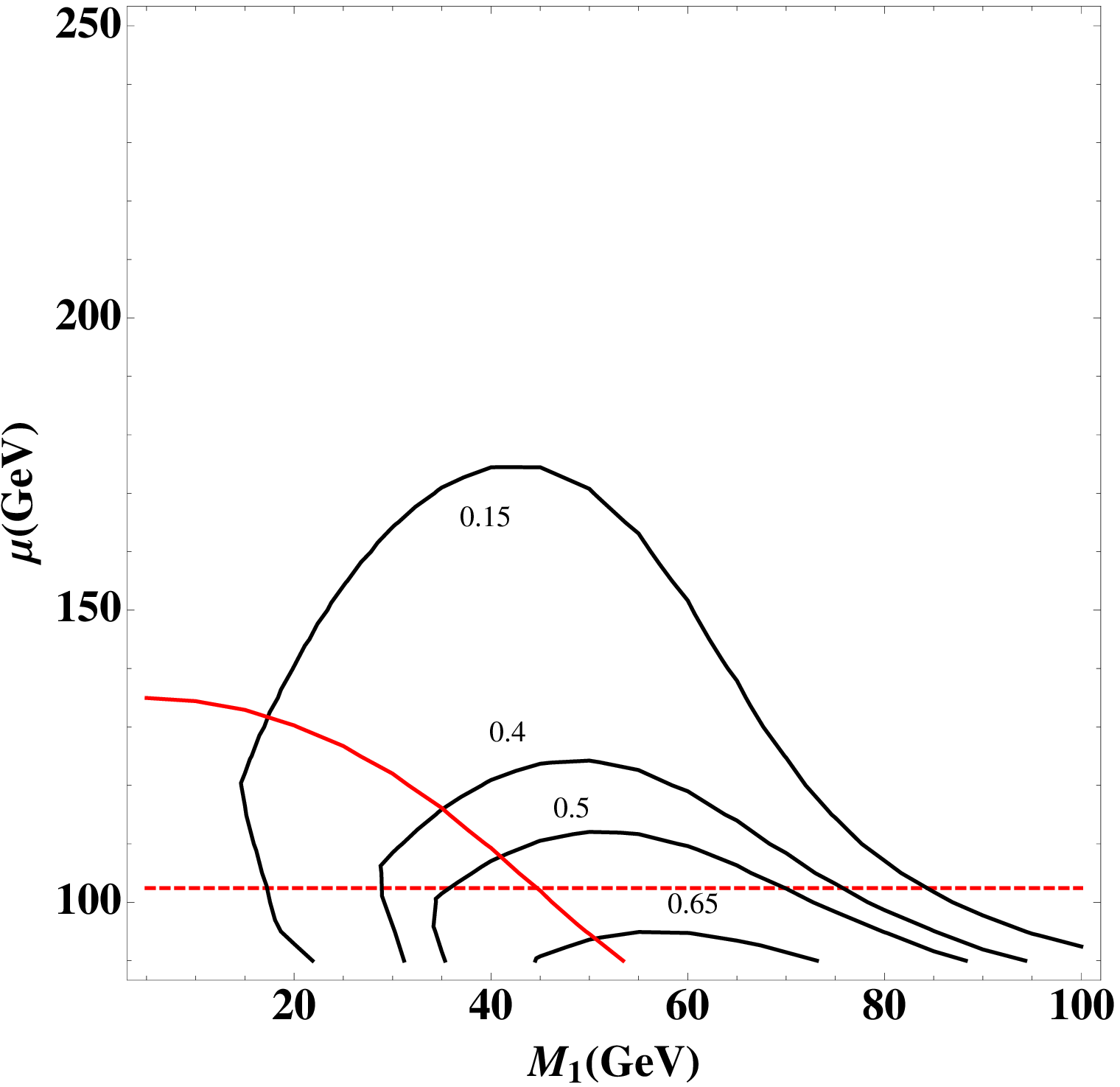}
\vspace*{0.3cm}
 \caption{\footnotesize 
Left: contours of 
${\rm BR} (h \rightarrow \chi^0_1 \chi^0_1) = 0.15,\; 0.4,\; 0.5, \; 0.65$ 
for $\tan\beta=10$ and $M_2=300$
 GeV. The area below the  thick (red) lines is excluded  by the $\Gamma_Z^{\rm inv}$  (solid)
and chargino mass (dashed) constraints. 
Right: same for $\tan\beta=40$.     }
\label{f1} 
\end{figure}

In the left plot of Fig.~\ref{f1}, we present our results in the $(M
_1, \mu)$ plane for $\tan\beta=10$ and $M_2=300$ GeV. The thick (red)
lines represent constraints from $\Gamma_Z^{\rm inv}$ (solid) and the
chargino mass (dashed) such that the area below them is excluded.
 For fixed $M_2$, the chargino constraint is a bound on $\mu$
  which only allows for values of $\mu$ above approximately 106 GeV.
  $\Gamma_Z^{\rm inv}$ exludes low $M_1$ and $\mu$ values, where the
  lightest neutralino has a substantial Higgsino component.  Given the
  constraints, we see that ${\rm BR_{inv}}$ can still be significantly above
  65\%. The shape of the constant ${\rm BR_{inv}}$ contours can be
  easily understood.  At low $M_1$, the Higgs decay into the lightest
  neutralinos is suppressed due to the small $N_{14}$. If $\mu$ is
  also relatively small, decays $h \rightarrow \chi_1^0 \chi_2^0$ and
  $h \rightarrow \chi_1^0 \chi_3^0$ become kinematically available,
  which reduces ${\rm BR} (h \rightarrow \chi_1^0 \chi_1^0)$ further
  and accounts for the kinks in the $\Gamma_Z^{\rm inv}$--excluded
  region.  ${\rm BR_{inv}}$ peaks at $M_1 \sim 30-60$ GeV, where
  $N_{14}$ is still significant and the kinematic suppression $\left(
    1- 4 m_{\chi^0_1 }^2/m_h^2 \right)^{3/2}$  has not yet set in.  In
  this range, $m_{\chi^0_1}$ varies between 20 and 50 GeV.  For $M_1 >
  80$ GeV, the invisible Higgs decay is strongly constrained by the
  chargino mass bound and becomes insignificant.  In summary, we find
  that if we take ${\rm BR_{inv}} < 40 \%$ as the bound, most of the
  region $\mu < 170$ GeV and $M_1 < 70$ GeV is disfavored by the
  invisible Higgs decay.

We conclude  that 
the Higgs decay bound is stronger than the $Z$--bound for intermediate
$M_1 \sim 30-70$ GeV. The reason for this is two--fold: (a) kinematic 
suppression of $Z \rightarrow \chi^0_1 \chi^0_1$ in this range, (b) linear
dependence of the Higgs--neutralino coupling on the Higgsino component
($N_{14}, N_{13}$) as opposed to the quadratic suppression in the $Z$ case.
Therefore, the Higgs decay covers a new territory, not explored by
other experiments.

It should be noted  that the massless neutralino scenario 
of \cite{Dreiner:2009ic}
is not excluded by these considerations. Choosing
\begin{equation}
M_1= \frac{ M_2 M_Z^2 \; \sin 2\beta \; \sin^2 \theta_W}
{\mu M_2 - M_Z^2 \; \sin 2\beta \; \cos^2 \theta_W} \;,
\end{equation}
one finds that $m_{\chi^0_1} =0$ at tree level.  
For values of $\mu$ allowed by the $\Gamma_Z^{\rm inv}$--bound, 
the massless
neutralino is mostly a bino and ${\rm BR} (h \rightarrow \chi_1^0
\chi_1^0)$ is typically around 10-20\% for $\tan\beta \sim 10$. 
A stronger experimental bound on ${\rm BR_{inv}}$ is necessary
to constrain this scenario.

\begin{figure}
\hspace*{0.1cm}
\includegraphics[width=6.9cm]{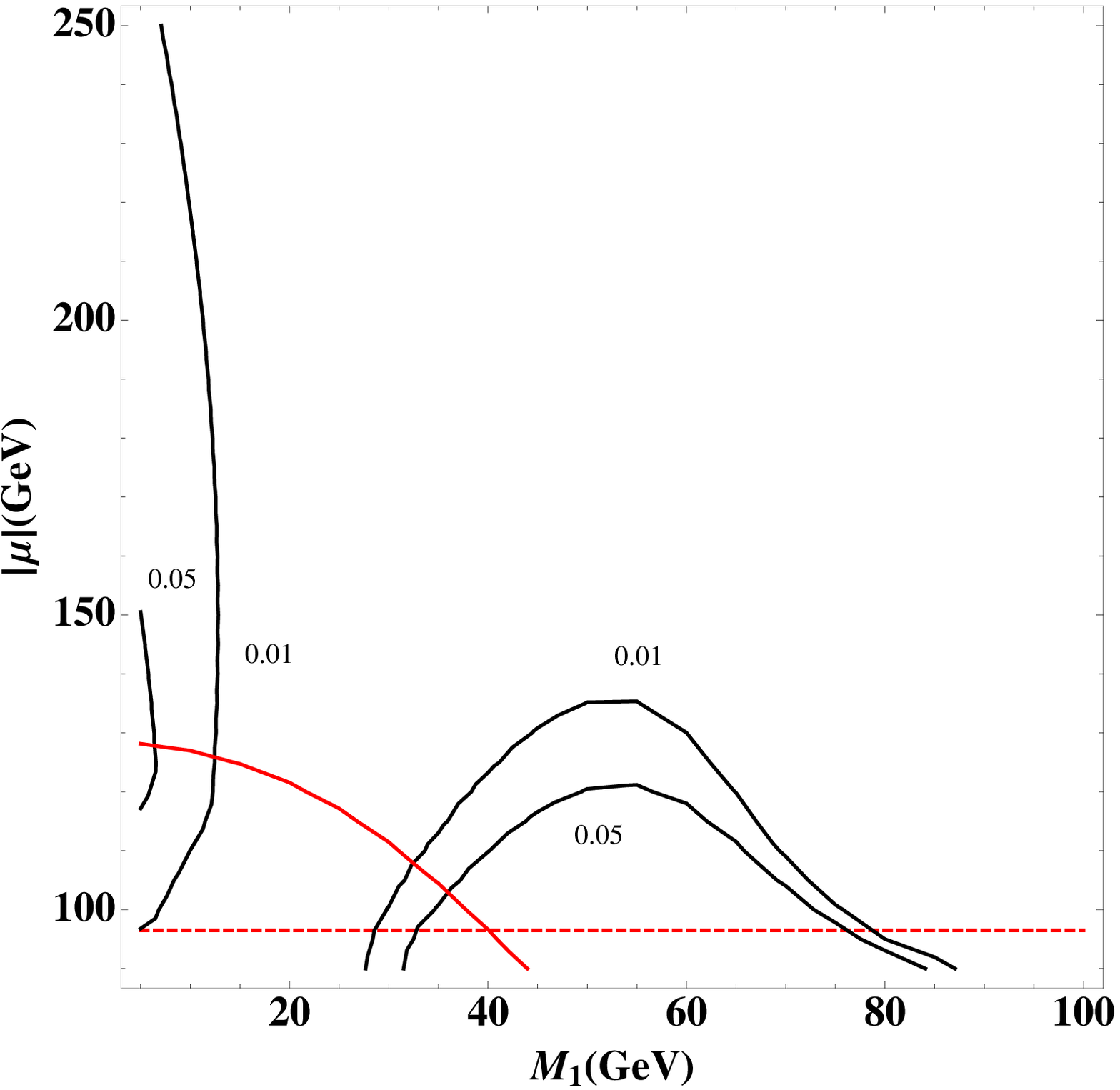}
\hspace*{0.1cm}
\includegraphics[width=8.5cm]{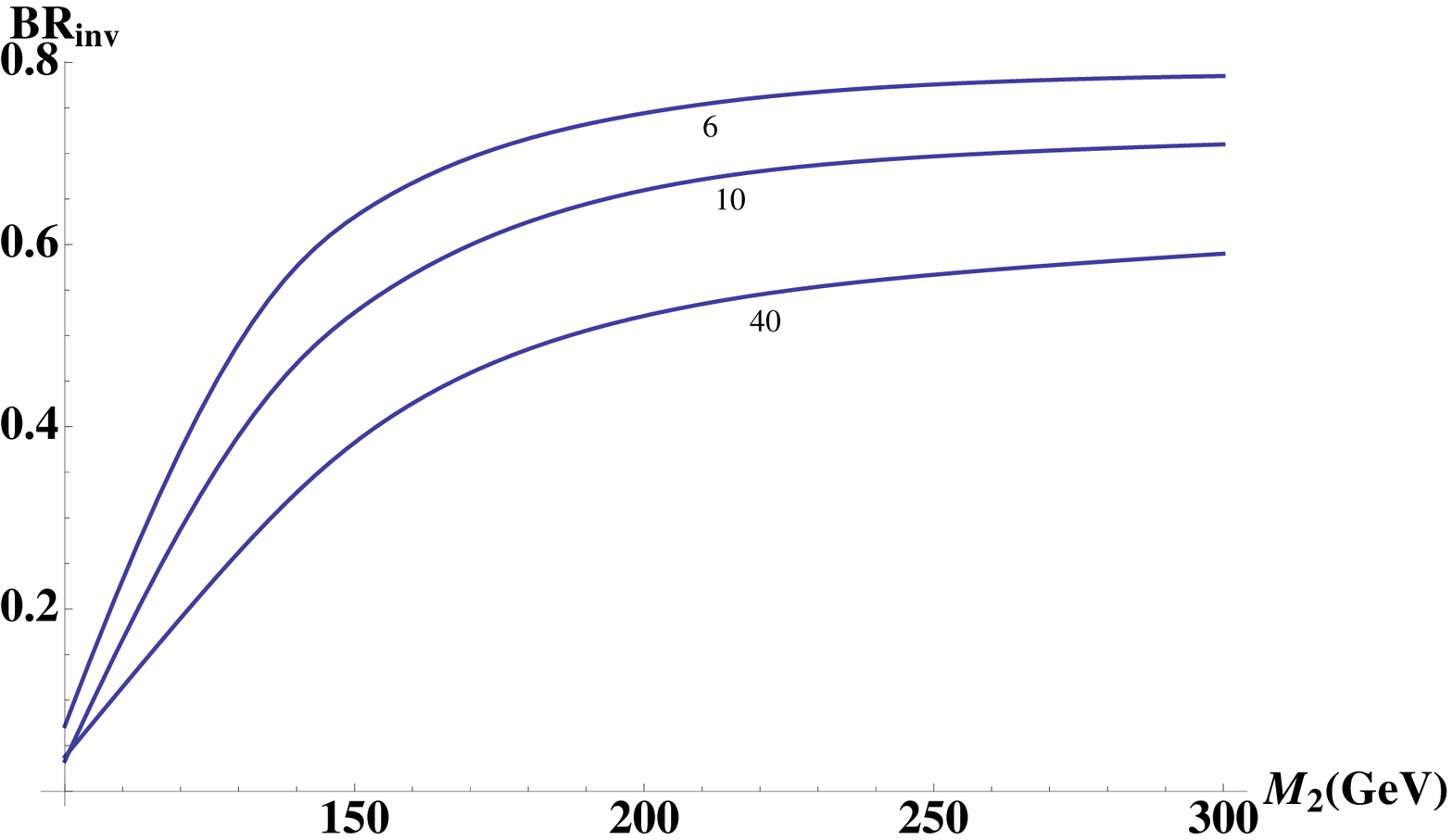}
\vspace*{0.3cm}
 \caption{\footnotesize 
Left: as in Fig.~\ref{f1} for $\tan\beta=10$ and $\mu <0$.
Right: maximal allowed  ${\rm BR_{\rm inv}}$ as a function of $M_2$ for 
$\tan\beta=6,\; 10,\;40$  (top to bottom).    }
\label{f2} 
\end{figure}

Below we summarize the dependence of ${\rm BR_{inv}}$ on the other parameters:

\begin{itemize}
\item $M_2$: lowering $M_2$  pushes up  the chargino bound on $\mu$ thus eliminating parameter 
space with the largest ${\rm BR_{inv}}$. 

\item $\tan\beta$: increasing $\tan\beta$ reduces the Higgs coupling to $\chi^0_1$, mostly 
due to the term $\cos\beta \; N_{13}$. As a result, ${\rm BR_{inv}}$ decreases.
For example, at $\tan\beta=40$, the disfavored region reduces to    $\mu < 120$ GeV and $M_1 < 70$ GeV
 (Fig.~\ref{f1}, right panel).

\item sign$\;\mu$: for $\mu <0$, the lighter chargino mass increases,
  relaxing the chargino bound.  On the other hand, the
  Higgs--neutralino coupling decreases due to a partial cancellation
  between $\sin\beta N_{14}$ and $\cos\beta N_{13}$. ${\rm BR_{\rm
      inv}}$ drops below 10-20\% (Fig.~\ref{f2}, left panel) imposing no
  significant constraint on parameter space.  Around $M_1 \sim 20$
  GeV, the cancellation is almost perfect and ${\rm BR_{\rm inv}}$ is
  negligible.
\end{itemize}

We thus find that ${\rm BR_{\rm inv}}$ imposes a significant constraint
on the neutralino sector of SUSY models, assuming that the Higgs 
signal gets confirmed. 
$h \rightarrow \chi^0_1 \chi^0_1$ can be the dominant Higgs decay channel 
with ${\rm BR_{\rm inv}}$ reaching 75\% for   moderate $\tan\beta$
and $M_2 > 200$ GeV (Fig.~\ref{f2}, right panel).
Values above 40\% are disfavored by the LHC Higgs signal which allows us
to place constraints on $\mu$ and $M_1$. These constraints are the strongest
for $\mu > 0$ and low $\tan\beta$, covering the $M_1$ values in the
kinematically allowed range for $h \rightarrow \chi^0_1 \chi^0_1$ up
to 80 GeV, and values of $\mu$ up to 200 GeV. 

It is clear that the constraints will get significantly stronger 
when the experimental limit on ${\rm BR_{\rm inv}}$ reaches a 10\% level.
For example, most of the parameter region shown in Fig.~\ref{f1} (left)
would be excluded. The massless neutralino scenario would also
be strongly constrained since the typical   ${\rm BR_{\rm inv}}$ 
is around 10-20\% in this case.
Further bounds on invisible Higgs decay can come  from monojet analyses
(see e.g. \cite{Djouadi:2012zc}), although their impact is expected to be less significant.

\section{Conclusion}

The tentative Higgs signal reported by the LHC collaborations appears
to agree well with the SM expectations.  In this paper, we have
studied implications of this observation for the neutralino sector of
SUSY models. The SM--like Higgs can decay into  a pair of the
  lightest neutralinos with the branching ratio up to 75\%. As
invisible Higgs decay is constrained by the existing data, we find
that most of the parameter region $\mu < 170$ GeV, $M_1 <70$ GeV at
$\tan\beta \sim 10$ and $\mu < 120$ GeV, $M_1 <70$ GeV at $\tan\beta
\sim 40$ is disfavored.

This conclusion depends only weakly on the other SUSY parameters. In particular,
the current bounds on superpartners suggest that the sfermion/gluino masses are
in the TeV range. It is therefore a good approximation to assume that the lightest
MSSM Higgs is very similar to the SM Higgs. The drastic difference however could 
appear in its invisible decays, if the decay into neutralinos is kinematically 
allowed. This allows us to set constraints on the Higgs--neutralino coupling,
which is controlled mostly by  $\mu$ and $M_1$. 
It is important to note that these constraints are ``direct'' in the sense that 
they do not rely on further assumptions
such as gaugino mass unification or specific SUSY  decay chains, unlike many previous 
analyses \cite{Nakamura:2010zzi}.

{\bf Acknowledgements.} The work of JSK is supported by the ARC Centre
of Excellence for Particle Physics at the Terascale. The work of HKD  was supported
by the BMBF Verbundprojekt HEP-Theorie under the contract 0509PDE.
JSK thanks A. Williams for reading the manuscript.

\end{document}